\documentclass[journal=jpclcd,manuscript=article]{achemso}

\usepackage[version=3]{mhchem} 
\usepackage{amsthm,amsmath,amsfonts,bm}

\newcommand{\tensor}[1]{\underline{\bm{#1}}}

\author{Massimiliano Comin}
\author{Simone Fratini}
\author{Xavier Blase}
\author{Gabriele D'Avino}
\email{gabriele.davino@neel.cnrs.fr}
\affiliation{Grenoble Alpes University, CNRS, Grenoble INP, Institut Néel, 25 rue des Martyrs, 38042 Grenoble, France}

\usepackage{color}

\title[]{
Doping-induced dielectric catastrophe  prompts free-carrier release in organic semiconductors }

\abbreviations{IR,NMR,UV}
\keywords{doping, organic semiconductors, dielectric screening, charge transfer, polarizability, QM/MM, dielectric catastrophe, insulator-conductor transition}

\begin{document}





\begin{abstract}
%
The control over material properties attainable through molecular doping is essential to many technological applications of organic semiconductors, such as OLED or thermoelectrics. 
These excitonic semiconductors typically reach the degenerate limit only at impurity concentrations of 5-10\%, a phenomenon that has been put in relation to the strong Coulomb binding between charge carriers and ionized dopants, and whose comprehension remained elusive so far. 
This study proposes a general mechanism for the release of carriers at finite doping in terms of collective screening phenomena. 
A multiscale model for the dielectric properties of doped organic semiconductor is set up by combining first principles and microelectrostatic calculations. 
Our results predict a large nonlinear enhancement of the dielectric constant (ten-fold at 8\% load) as the system approaches a dielectric instability (catastrophe) upon increasing doping.
This can be attributed to the presence of highly polarizable host-dopant complexes, plus a nontrivial leading contribution from dipolar interactions in the disordered and heterogeneous system. 
The enhanced screening in the material drastically reduces the (free) energy barriers for electron-hole separation, rationalizing the possibility for thermal charge release. The proposed mechanism is consistent with conductivity data and sets the basis for achieving higher conductivities at lower doping loads. 
\end{abstract}

The ability to control the charge carrier density and transport levels is key to the success of semiconducting technologies and organic semiconductors (OSCs) made of $\pi$-conjugated molecules or polymers make no exception.
Chemical doping by means of electron attracting or withdrawing impurity molecules emerged as a successful approach that lies at the foundation of stable and efficient organic optoelectronic and thermoelectric devices.\cite{Wal07,Jac17,Sal16,Zhao20}
While recent years have witnessed remarkable advances concerning the chemistry\cite{Lin17,Yam19,Yur19} and the structural control\cite{Kan16,Hase18,Jac18,War19} 
of doped OSCs, 
the microscopic phenomena that govern
molecular doping remain to date unclear.

A central open question concerns the release of free doping-induced charge carriers in low-dielectric constant organic materials.
Experimental data for small-molecules and polymers show that a 
boost of the electrical conductivity by orders of magnitude is typically achieved at doping loads of about 5-10\%, i.e. at concentrations that are orders of magnitude higher than in inorganic semiconductors.\cite{Kle12,Men15,Pin13}
Several studies have shown that the conductivity follows a thermally activated behavior,\cite{Mae01,She03,Olt12,Schwa19}
and the  activation energy has been related to the Coulomb interaction between an ionized dopant and the charge injected in the semiconductor.\cite{Schwa19}
Photoemission measurements \cite{Gau18} and theoretical calculations\cite{Li17,Li19,Pri20} set this binding-energy in the {400-700 meV} range in the low-doping regime, a value that is too large to permit a significant release of free carriers at room temperature.

Experimental data for $p$ and $n$-doped OSCs have shown a universal tendency for a conspicuous reduction of the activation energy with dopant concentration.\cite{Schwa19,She03,Olt12,Zha18}
Besides, the conductivities of a set of polymers heavily-doped by ion-exchange have been shown to be independent on the ion size,\cite{Jac21} which is at odds with the interpretation of transport limited by Coulomb interactions.
This set of evidences, together with the high loads needed to boost conductivity, points to a collective depinning mechanism for charge carriers taking place upon increasing doping concentration, beyond trap filling effects at ultralow doping.\cite{Olt12,Tie15}

Theoretical studies mostly focused on the infinite dilution limit, discussing single host-dopant complexes with quantum or hybrid quantum/classical (QM/MM) methods.\cite{Sal12,Men15,Gau18,Val19,Pri20} Some of us applied embedded many-body perturbation techniques and an essential-state model for charge transfer (CT) degrees of freedom to the paradigmatic case of F4TCNQ-doped pentacene,\cite{Ha09,Kle12,Sal12} showing that the full dopant ionization takes place at room-temperature thanks to the excitonic e-h interaction and structural relaxation.
That study  
predicted also the occurrence of very low-energy (down to 34 meV) CT excitations in spite of acceptor levels being very deep in the host gap.
Coming to transport, Kinetic Monte Carlo simulations proved able to reproduce the doping-induced conductivity enhancements observed in experiments.\cite{Tie18,Fed19,Fed20,Koo20}
Those results have been interpreted in terms of the favorable effect of the energetic disorder, sourced from the Coulomb field of doping-induced charges, assisting  
charge carriers' release.

In this article, we investigate the effect of collective screening phenomena on the release of doping-induced charges.
Following a multiscale approach, we first evaluate \textit{ab initio} the polarizability associated with low-energy CT degrees of freedom in host-dopant complexes, showing that this
contribution is 
as large as ten times that of a single molecule.
This information is then used to build a microscopic model for doped OSCs that allows us to obtain the dielectric properties of doped OSCs as a function of the impurity concentration.
Our calculations reveal a large enhancement of the relative permittivity 
$\varepsilon_\mathrm{r}=\varepsilon/\varepsilon_0$ upon doping, with an order of magnitude increase at 8\% concentration,  which implies a drastic suppression of Coulomb energy barriers for the release of free charge carriers. 
The origin of this enhancement is rooted in the collective response of highly-polarizable host-dopant complexes 
dopants in a disordered system, whose susceptibility increases upon doping and diverges 
at the approach of the dielectric catastrophe.



\begin{figure}[ht]
\includegraphics[width=\textwidth]{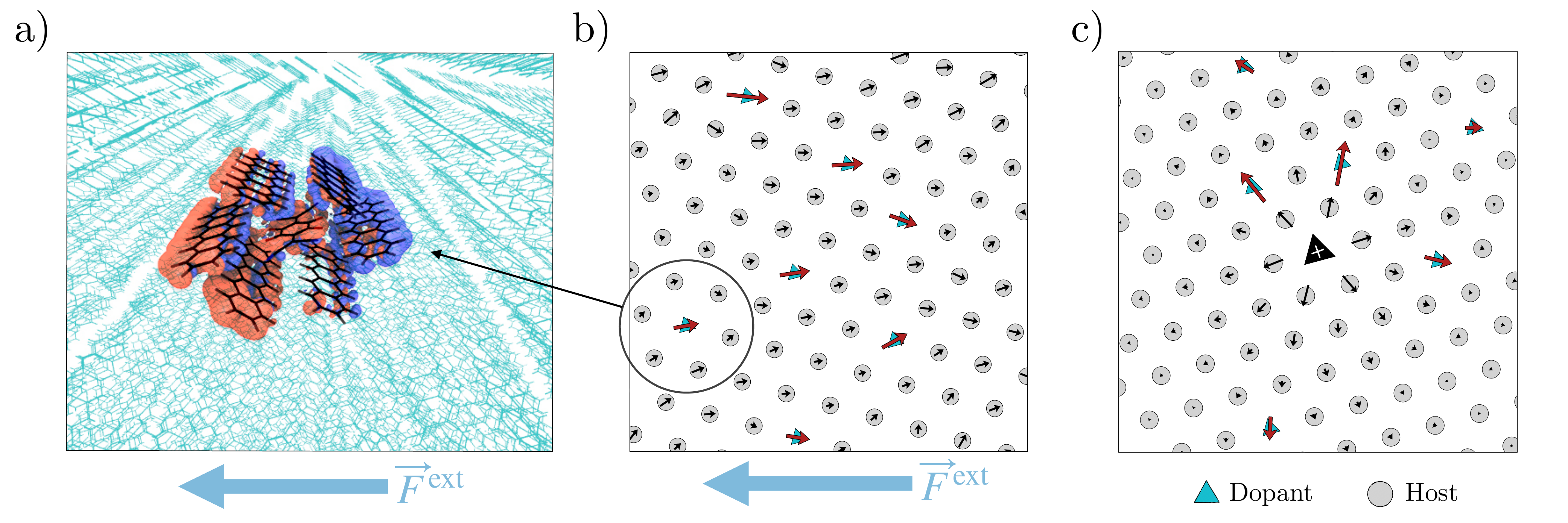}
\caption{Illustration of the proposed multiscale approach to the dielectric properties of doped organic semiconductors.
(a) QM/MM calculation of a F4TCNQ-pentacene complex (QM region) in the host pentacene crystal (MM, cyan wireframe representation). 
The red/blue surface depicts the electron density induced by an electric field, ${\Delta}n(F^\mathrm{ext})=n(F^\mathrm{ext})-n(0)$, with isocontour drawn at negative/positive value. 
The large polarizability of the complex is determined by such a displacement of the electron density within the QM region.
%
Sketch of the dipoles induced in a doped semiconductor ($\rho=8\%$) (b) by a homogeneous field and (c) by a point charge. 
Host and dopant molecules are shown as circles and triangles, respectively, dipoles of CT complexes are drawn in red.
The dipoles of host and dopant sites are not in scale.
}
\label{f:qmmm}
\end{figure}

Our analysis starts from first-principles calculations of the polarizability of host-dopant CT complexes and considers the paradigmatic case of a F4TCNQ impurity in the pentacene crystal, see Fig.~\ref{f:qmmm}a.
This system is described with hybrid quantum/classical (QM/MM) calculations, where we treat the complex composed by the dopant and its first shell of six pentacene neighbors at the density functional theory (DFT) level, employing a hybrid functional tuned to reproduce the gap obtained from accurate embedded $GW$ calculations.\cite{Li17}
The surrounding crystalline environment is described by an atomistic polarizable model.\cite{Li18,Dav14}

%

Our results in Table~\ref{t:alpha}
reveal that the polarizability of an isolated (gas-phase) complex is significantly larger than the sum of the molecular polarizabilities, testifying a contribution from intermolecular CT interactions. 
Moreover, the polarizability of the complex roughly doubles when accounting for the MM environment, an observation that can be equivalently explained in terms of the screening of the dipolar field generated by the induced charge density (shown in Figure~\ref{f:qmmm}a), or as a result of the stabilization of CT excitations by the polarizable environment.
The full QM/MM polarizability of the complex, $\alpha_\mathrm{cpx}=804$~\AA$^3$, accounts for both an intramolecular an intermolecular CT contribution.
Upon dissecting the two terms with an approximate subtractive scheme, we have been able to quantify a leading CT contribution $\alpha_\mathrm{CT}=534$~\AA$^3$, which is above ten times that of a single pentacene molecule.
Full detail on polarizability calculations are provided as Supporting Information (SI).

\begin{table}[]
  \begin{tabular}{lll}
  \hline
    System  & Method & $\alpha$ (\AA$^3$)  \\
    \hline
    Pentacene   & DFT gas & 47 \\
    F4TCNQ            & DFT gas & 33 \\
    F4TCNQ-pentacene  & DFT gas & 401 \\
    F4TCNQ-pentacene  & DFT/MM  & 804 \\
    F4TCNQ-pentacene, CT & DFT/MM  & 534 \\
    F4TCNQ-pentacene, CT & CT model  & 357 \\
\hline
\end{tabular}
\caption{Polarizability of pentacene and F4TCNQ molecules and their complex shown in Figure~\ref{f:qmmm}a, calculated at DFT level in the gas and in the solid state (QM/MM embedding). 
The last two lines report the polarizability associated to intermolecular CT degrees of freedom, calculated at QM/MM level and with a generalized Mulliken model for intermolecular CT. 
Note that the CT complex polarizability is one order of magnitude larger than that of a single molecule.  }
\label{t:alpha}
\end{table}

Such a remarkably large value compares well with the estimate from a simpler Mulliken-like model for intermolecular CT (see Table~\ref{t:alpha}), which has been carefully parameterized for F4TCNQ-doped pentacene and validated against embedded Bethe-Salpeter Equation calculations.\cite{Li17}
As shown in detail in the SI, model calculations allowed us to extend and generalize our first-principles findings to other systems, including a realistic amorphous sample of F6TCNNQ-doped NPB.\cite{Li19}
The CT polarizabilities of complexes extracted from this amorphous system present an order of magnitude variability, with average of 577~\AA$^3$ and values as large as 1000~\AA$^3$.
Furthermore, model calculations spanning the set of microscopic parameters of relevance for doped OSCs allowed us to conclude on very general grounds that host-dopant complexes featuring a partial or a full CT in the ground state are intrinsically highly polarizable as a result of the presence of dipole-allowed low energy excitations.

Having characterized the polarizabilities associated with host-dopant CT complexes, we are in the position to set up a model for the dielectric properties of OSCs at finite doping load.
As a first step towards a more realistic description, we describe this heterogeneous dielectric medium with an induced-dipole scheme on a face-centered cubic (FCC) lattice, whose sites represent either a host molecule or a dopant complex, see Fig.~\ref{f:qmmm}b,c.
Dopants are randomly distributed over the lattice and a polarizability $\alpha_\mathrm{CT}=500$~\AA$^3$
is assigned to these sites. 
The polarizability of host  sites is set to $\alpha_\mathrm{host}=50$~\AA$^3$,
which corresponds to that of a pentacene molecule.
This, for the lattice constant of 12.26 \AA\ used throughout this work, corresponds to a dielectric constant of $\varepsilon_\mathrm{r}=3.5$ for the pristine host, according to the standard Clausius-Mossotti relation.
We note that while the dielectric properties and the related dipole-field sums have been extensively investigated in high-symmetry lattices of equivalent sites,\cite{mueller_PhysRev35,Colpa_Physica71A,Purvis_taylor_PRB82,Van15} to the best of our knowledge, the effect of disorder and inhomogeneity remains completely unexplored to date.


Following the common treatments of dipolar linear response, the dipoles induced in the system by the electric field of permanent sources $\mathbf{F}_0$ (e.g. an external field or a point charge inside the material)  can be obtained by solving the linear system
\begin{equation} \label{e:lineq}
    \tensor{H}\bm{\mu} = \bm{F}_0
\end{equation}
where $\bm{\mu}$ and  $\bm{F}_0$ are vectors of the Cartesian components of induced dipoles and permanent fields at the lattice sites. 
The Hessian matrix $\tensor{H}$ consists of the second derivatives of the system's energy with respect to the interacting dipoles, accounting for site polarizability and dipole-dipole interactions.
The advantage of this approach consists in the possibility to describe with numerically exact calculations inhomogeneous systems of thousands of molecular sites, possibly accounting for periodic boundary conditions: as a result, the present method allows to span the interval of doping loads of relevance for experiments. 

A first insight on the effect of the increasing dopant density, $\rho$, can be obtained from the spectrum of the Hessian matrix of an extended system.
The Hessian eigenvalues $\lambda$ quantify the stiffness of the collective polarization modes of the system of interacting polarizabilities, i.e. their inverse, $\lambda^{-1}$, is proportional to the polarizability associated to each normal mode. 
In particular, vanishing positive Hessian eigenvalues determine a finite polarization even in the presence of a tiny perturbation, while modes characterized by $\lambda<0$ are unstable, i.e. develop a spontaneous polarization even in the absence of an electric field.


\begin{figure}[ht]
\includegraphics[width=0.8\textwidth]{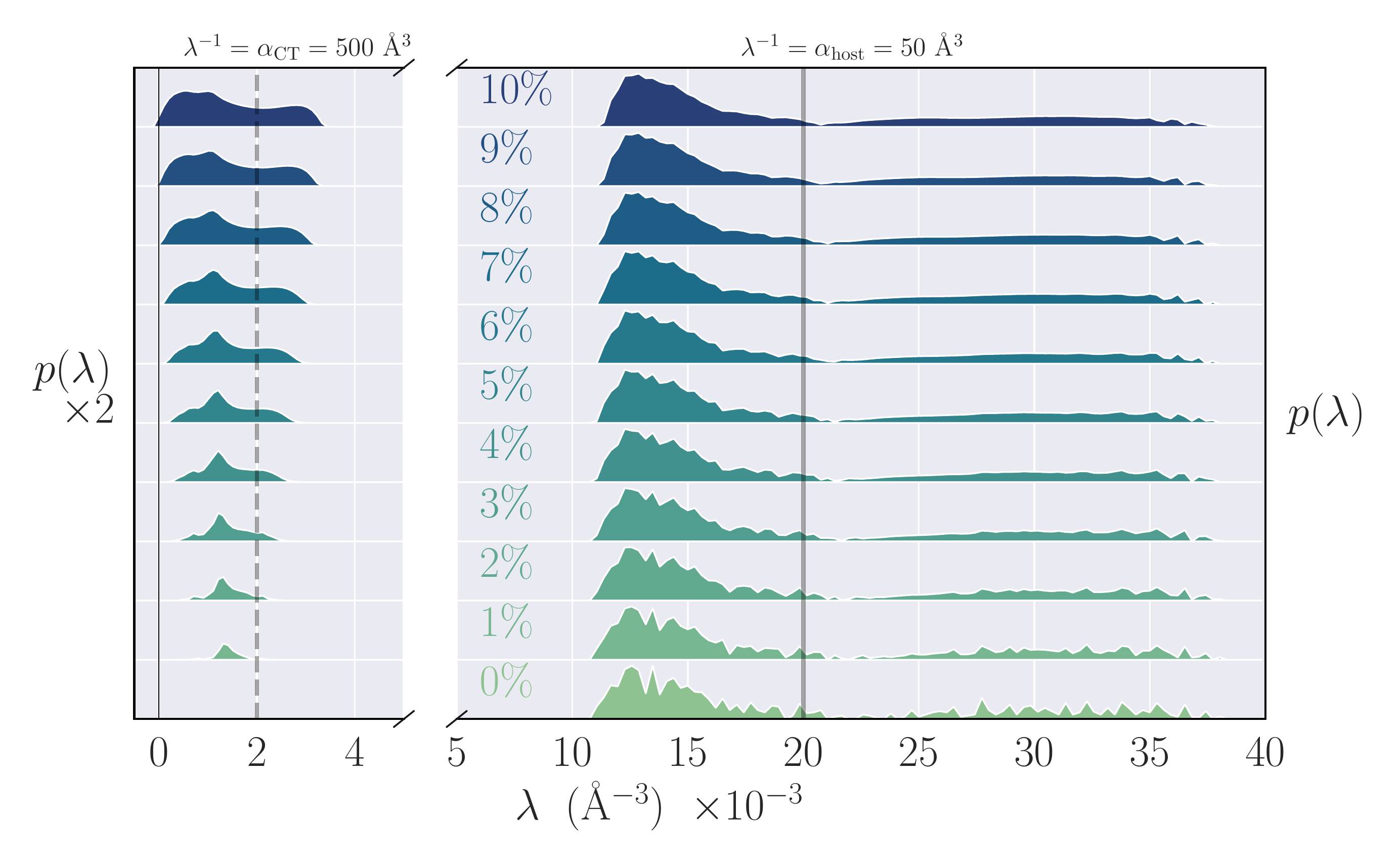}
\caption{
Distributions of the Hessian matrix eigenvalues, measuring the stiffness of the collective polarization modes in a doped OSC, for increasing impurity concentration $\rho$.
Results obtained with an induced-dipole model on a lattice, with site polarizabilities corresponding to the typical values for doped OSCs, e.g. pentacene-F4TCNQ (see Table~\ref{t:alpha}).
The vertical lines mark the inverse polarizabilities of non-interacting host molecules and host-dopant CT complexes. 
The distributions are characterized by two main features associated to the modes of host sites 
($\lambda>10\cdot 10^{-3}$~\AA$^{-3}$) 
and to CT complexes ($\lambda<4\cdot 10^{-3}$~\AA$^{-3}$).
The latter features an increasing dispersion with $\rho$, with the low-$\lambda$ tail of the distribution approaching, and eventually surpassing, zero.
This interaction-driven softening results is an enhanced susceptibility to perturbations.
The y-scale axis in the left panel has been magnified for better visualization.
}
\label{f:dos}
\end{figure}

The distribution of $\lambda$ values at different doping loads is shown in Figure~\ref{f:dos}, where the vertical lines mark $\alpha_\mathrm{host}^{-1}$ and $\alpha_\mathrm{CT}^{-1}$, i.e. the stiffness of non-interacting host and dopant sites, respectively.
A sizable eigenvalue dispersion due to inter-site interactions is obtained already in the pristine system, $\rho$=0\%.
This distribution is bimodal, with a peak at $13 \cdot 10^{-3}$~\AA$^{-3}$ (transverse polarization modes branch), and a much broader feature centered around $30 \cdot 10^{-3}$~\AA$^{-3}$ (longitudinal modes).
Upon increasing doping, the shape of the distribution barely changes in the region corresponding to the host sites ($\lambda> 10 \cdot 10^{-3}$~\AA$^{-3}$).
On the other hand, new peaks resulting from highly polarizable dopant-host complexes appear around $\alpha_\mathrm{CT}^{-1}$ (dashed vertical line).
These peaks grow in intensity and significantly broaden with increasing concentration as a result of the interaction between complexes.
Most interestingly, the low-$\lambda$ tail of the distribution approaches zero upon increasing $\rho$, with a 1.6\% fraction of the modes of the dopant complexes band becoming unstable ($\lambda < 0$) at 8\% doping.
This softening of the collective polarization modes with doping signals an enhancement of the susceptibility as the system moves toward a \emph{dielectric catastrophe}.\cite{Her27}

We next turn our attention to the bulk dielectric constant, that we compute by generalizing the approach described in Refs.~\cite{Tsi01,Dav16} to large inhomogeneous systems.
As sketched in Figure~\ref{f:qmmm}b, we determine the polarization $\mathbf{P}$ induced by a homogeneous electric field $\mathbf{F}_0=\mathbf{F}_\mathrm{ext}$ applied along the three Cartesian directions.
The dielectric tensor is then obtained upon applying the appropriate depolarization correction as
\begin{equation} \label{e:eps_r}
 \tensor{\varepsilon}_\mathrm{r} = 
 1 + (1 - \tensor{\zeta}/3 )^{-1} \cdot \tensor{\zeta},
\end{equation}
where 
$\tensor{\zeta}$ is the susceptibility to the external field, i.e. $\mathbf{P}=\varepsilon_0\tensor{\zeta}\mathbf{F}_\mathrm{ext}$.
Because the response of the system averaged over the realizations of the dopants' positional disorder is isotropic,  in the following we will only report the scalar dielectric constant.

\begin{figure}[ht]
\includegraphics[width=\textwidth]{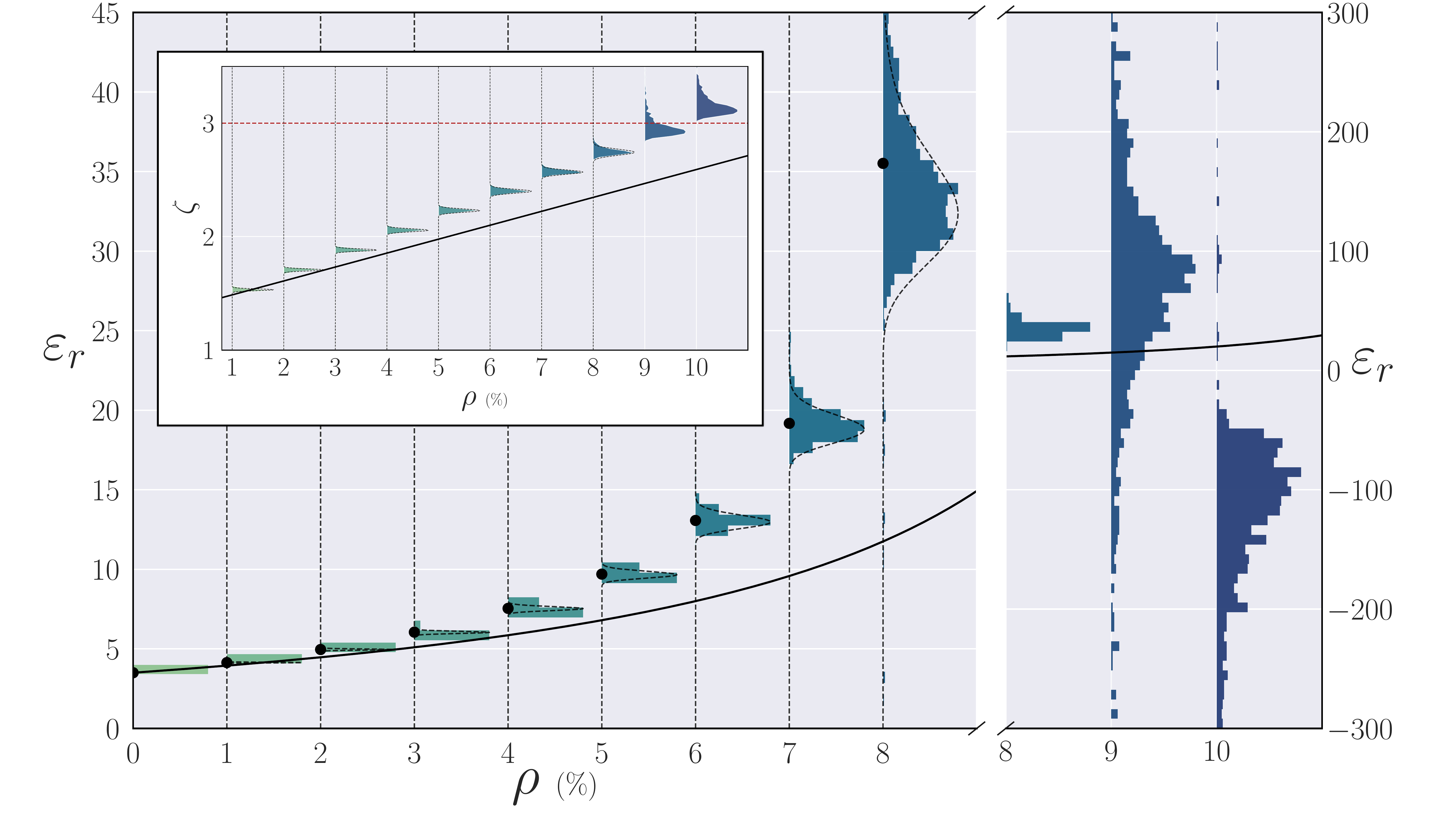}
\caption{
Enhancement of the bulk dielectric constant of a doped organic semiconductor with the impurity concentration $\rho$.
Results obtained with an induced-dipole model on a lattice, with site polarizabilities corresponding to the typical values for doped OSCs, e.g. pentacene-F4TCNQ (see Table~\ref{t:alpha}).
The main panel shows the distribution (histograms) and harmonic mean (dots) of $\varepsilon_\mathrm{r}$ obtained upon sampling over dopants positional disorder.
The full line corresponds to the dielectric constant obtained for an effective homogeneous medium with mean site polarizability $\alpha_\mathrm{avg}(\rho)$, see text.
The inset shows the dependence of the external-field susceptibility $\zeta$ on doping, displaying numerical results for the inhomogeneous lattice (distributions and their Gaussian fits as dashed lines) and $\zeta(\rho)$ for an effective homogeneous medium (full line). 
The Gaussian fits are then analytically transformed into distributions for $\varepsilon_\mathrm{r}$, shown as dashed lines in the main panel.
The right-hand panel illustrates the system behavior at the dielectric catastrophe, note the different scale on the ordinate.
}
\label{f:eps_bulk}
\end{figure}

Our calculations predict a divergent behavior of the dielectric constant, leading to a large non-linear enhancement as function of the doping load, shown by the black dots in Figure~\ref{f:eps_bulk}.
This amounts to a four (ten) fold increase of $\varepsilon_\mathrm{r}$ at 6\% (8\%) doping with respect to the pristine OSC, which can have a dramatic effect on the release of Coulombically bound charges, as discussed later.
We emphasize that such a striking increase of $\varepsilon_\mathrm{r}$ occurs at the typical concentrations leading to a boost in the measured electrical conductivity.\cite{Wal07,Sal16,Jac17}

The  enhancement in the dielectric constant can be rationalized by  looking at the polarization response to the external field, $\zeta$.
The distributions of $\zeta$ values obtained upon sampling the dopants' positional disorder are shown as an inset in Fig.~\ref{f:eps_bulk}.
These distributions are symmetric, nicely approximated by Gaussians (black dotted lines) up to $\rho=8$\%, and develop some skewness at larger doping, i.e. when the dielectric becomes more and more unstable.
According to Eq.~\ref{e:eps_r}, the dielectric constant diverges (dielectric catastrophe) at $\zeta=3$ (horizontal dashed line). 
This  determines a remarkable, non-linear amplification of $\varepsilon_\mathrm{r}$ upon approaching the singularity from $\zeta<3$, and negative values afterwards, the latter marking the instability. 
This explains the histogram distributions of $\varepsilon_\mathrm{r}$, shown in the main panel of Figure~\ref{f:eps_bulk}, which shift to higher values, broaden and become more and more skewed upon approaching the dielectric catastrophe by increasing $\rho$. 
This behaviour is captured to a good approximation by evaluating analytically the distribution of $\varepsilon_\mathrm{r}(\zeta)$ (dashed lines, main panel) via Eq.~\ref{e:eps_r}, using the Gaussian fits of the $\zeta$ distributions.


The picture that emerges from our results is that two phenomena
cooperate in determining the boost of the dielectric constant with $\rho$.
The first one is the straightforward effect of  increasing the concentration of highly polarizable CT complexes. 
This effect can be captured by considering an effective medium with average site polarizability $\alpha_\mathrm{avg}(\rho)=(1-\rho) \alpha_\mathrm{host} + \rho \alpha_\mathrm{CT}$. 
This effective medium approximation, shown as a solid line in Figure~\ref{f:eps_bulk}, only accounts for $26\%$  of the doping-induced enhancement of $\varepsilon_\mathrm{r}(\rho=8\%)$ (dots).
The remaining, dominant, contribution originates from cooperative dipolar interactions taking place in the disordered lattice, which result in microscopic dipole fields that are strongly polarizing (i.e. enhancing the external field) at highly-polarizable dopant sites (see SI, Figure S3).
As a result, the external-field susceptibility $\zeta$ presents a faster increase with doping in the inhomogeneous than in the effective homogeneous medium (inset of Figure~\ref{f:eps_bulk}, full line), attaining the divergence at a lower doping.
This marks a crucial qualitative difference between a homogeneous dielectric, which obeys the Clausius-Mossotti equation, and an intrinsically chemically inhomogeneous disordered medium, such as a doped OSC.

\begin{figure}[p!]
\includegraphics[width=\textwidth]{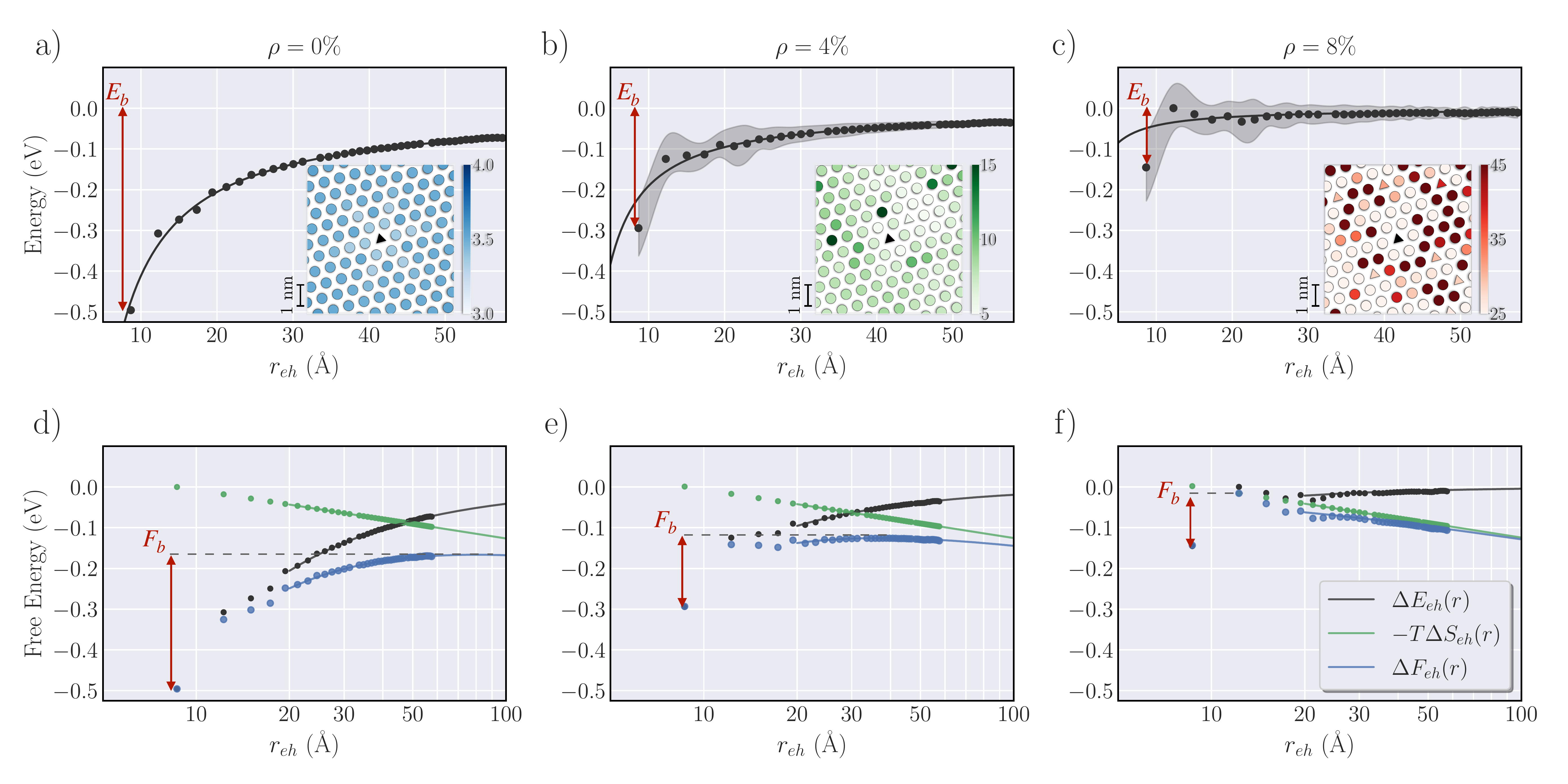}
\caption{
Energy and free-energy profiles for charge separation, showing the large reduction of the barrier for free carrier release ($E_b$ and $F_b$) upon increasing the doping load, as a result of dielectric screening and entropy. 
(a-c) Energy profiles, mean value (dots) and standard deviation (shaded areas) for the different coordination shells sampled over dopants disorder. 
The full line corresponds to a screened Coulomb potential calculated with bulk $\varepsilon_\mathrm{r}$.
The insets show the non-local dielectric constant $\varepsilon_\mathrm{r}(\mathbf{r},\mathbf{r}')$, computed for a source charge at $\mathbf{r}'$ (central site) for a given realization of dopants position.
Circles and triangles represent host and dopant sites, respectively.
(d-e) Room-temperature free energy profiles, partitioned into energetic and entropic contributions (semi-logarithmic scale). 
Dots correspond to numerical results, lines are analytic expressions for the continuum.
}
\label{f:Vr}
\end{figure}

Finally, we discuss the energetics of charge dissociation in the presence of the enhanced dielectric screening sourced from highly-polarizable host-dopant complexes.
To such an aim, we calculate the collective response to the field of a point-charge $\mathbf{F}_0=\mathbf{F}_\mathrm{q}$ (see Figure~\ref{f:qmmm}c), which gives us access to charge separation energy profiles as a function of the electron-hole distance $r_{eh}$, shown in panels a-c of Figure~\ref{f:Vr}.
In the infinite dilution limit ($\rho=0$, panel a), except for small oscillations at short distance, the energy profile (dots) has the form of a Coulomb potential, screened by the bulk $\varepsilon_\mathrm{r}=3.5$
(full line), similar to  analogous atomistic results.\cite{Li17} 
The energy barrier to free charges sitting on nearest-neighbour molecules ($E_b$  in Figure~\ref{f:Vr}a-c) amounts to 0.5 eV, consistent with first-principles calculations \cite{Li17,Gau18} and conductivity data at low doping.\cite{Schwa19}

The situation radically changes at finite dopant concentration.
First, the screening provided by dopant complexes determines 
a sizable suppression of the average charge-separation energy barrier, reducing to 0.29 and 0.15 eV at 4 and 8\% doping, respectively. 
We anticipate that this large screening effect provides 
a decisive contribution to the charge release mechanism explained below.
Second, our calculations provide insights into the effects of the inhomogeneity of the medium.
This determines a substantial spread in the energy profile for charge separation (gray shaded areas in Figure~\ref{f:Vr}b,c), with important local deviations from the screened Coulomb potential with bulk $\varepsilon_\mathrm{r}$, which is recovered only at large $r_{eh}$. 
The spatial inhomogeneity of the system can be best appreciated from the fluctuations of the non-local microscopic dielectric constants $\varepsilon_\mathrm{r}(\mathbf{r},\mathbf{r}')$, shown in the insets.
These fluctuations, which would be missed by reasoning only in terms of the bulk $\varepsilon_\mathrm{r}$, may determine pathways for charge separation that are energetically more favorable than the average one.

In order to ultimately address the question of charge carriers release at room temperature, we now assess the free energy profile for charge separation, shown in Figure~\ref{f:Vr}d-f.
The free energy accounts for the entropy contribution resulting from the radial density of microstates, increasing as $r_{eh}^2$ in three dimensions.\cite{Gregg11}
The entropy variation always assists charge separation, leading to a further decrease of the free-energy barrier ($F_b$ in Figure~\ref{f:Vr}d-f), which reduces to 0.17 and  0.13 eV at 4 and 8\% doping,
rationalizing the possibility for the thermal release of free carriers at the impurity concentrations that actually correspond to the conductivity boosts seen in experiments.


In summary, we have proposed a model description of collective screening phenomena in doped organic semiconductors and their implications on the release of Coulombically bound charges introduced upon doping.
Our treatment based on local polarizabilities associated to  highly-polarizable host-dopant CT complexes and host molecules, which strictly holds upon approaching conducting states from the insulating side (i.e. for bound charges), predicts a factor ten increase of the bulk dielectric constant at doping loads of 8\% and very large fluctuations at the microscopic scale.
This enhanced screening, together with entropic effects, drastically reduces the Coulomb barrier for charge separation, providing a mechanistic explanation for the impurity concentration dependence of the activation energies extracted from conductivity data.\cite{Schwa19}

Divergent dielectric responses are \emph{de facto} expected in conducting states and arguments similar to ours, i.e. linking the polarizability of hydrogen-like impurities to $\varepsilon_\mathrm{r}$, have been proposed for doped silicon.\cite{Cas51,Dhar85,Mott85} 
We remark that the microscopic treatment proposed herein goes far beyond early works based on local-field corrections,\cite{Cas51,Dhar85} as it is able to capture the major role played by inhomogeneity and disorder at the molecular scale.

The present analysis builds a fresh picture of the doping-induced insulator-to-conductor transition in terms of an incipient  dielectric catastrophe, that is broadly consistent with available experimental data for organic semiconductors, including the 5-10\% loads needed to boost conductivity.\cite{Wal07,Sal16,Jac17}
The proposed mechanism is in apparent contradiction with the prevailing picture of a charge release prompted by energetic disorder due to the dipolar fields of host-dopant complexes, which was suggested by kinetic Monte Carlo simulations that correctly capture conductivity enhancements upon doping.\cite{Tie18,Fed19,Fed20,Koo20}
As a possibility to reconcile these two scenarios, we conjecture that the many-body screening phenomena we describe in the present work may be captured by kinetic Monte Carlo, although their observation requires to go beyond 
macroscopic observables such as conductivity or time and space averaged densities of states.
The latter cannot disclose any possible evidence for correlated charge motions at the origin of screening phenomena, whose only manifestation would be a spread in molecular site energies, i.e. energetic disorder.

We are confident this work will stimulate experimental studies targeting the observation of low-energy electronic excitations responsible for the large polarizability of dopant complexes, as well as the direct measurement of dielectric constant of doped organic semiconductors.
A better understanding and management of the dielectric properties of doped organic semiconductors represents a promising gateway to the achievement of higher conductivities at lower doping concentration.

\section{Methods}

QM/MM calculations were performed at the DFT PBEh($a$=0.4)/cc-pVTZ  level with the atomistic polarizable embedding scheme described in Ref.~\cite{Li18}.
The DFT/MM polarizability is calculated as numerical derivatives of the QM-region dipole induced by an electric field, accounting for the effect of the polarizable MM environment.
QM/MM calculations have been performed with the ORCA v4.2 and MESCal\cite{Dav14} codes.

The induced-dipole model for doped OSCs considered a FCC lattice, 
with dopants randomly distributed over sites with the constraint of not having first and second nearest neighbor impurities.
Results have  been obtained upon sampling over 1000 random realizations of the distribution of dopants over lattice sites (dopants' positional disorder).
The latter actually bear the polarizability of the host-dopant CT complex, hence the interaction between its polarizability and that of the first host neighbors has been neglected.
The linear system in Eq.~\ref{e:lineq} has been solved by full diagonalization or by matrix inversion (Cholesky factorization).
Bulk dielectric constant calculations were performed accounting for periodic boundary conditions in three dimensions.
Charge-separation radial energy profiles have been calculated by placing a source charge at $\mathbf{r}'$, and computing the energy of a probe charge $E(\mathbf{r})=eW(\mathbf{r},\mathbf{r}')$, where $W$ is the electrostatic potential screened by induced dipoles.
These calculations have been performed in open-boundary conditions, carefully extrapolating $W$ in the infinite system limit.
The nonlocal dielectric constant is obtained as 
$\varepsilon_\mathrm{r}(\mathbf{r},\mathbf{r}')=v(\mathbf{r},\mathbf{r}')/W(\mathbf{r},\mathbf{r}')$, where $v(\mathbf{r},\mathbf{r}')=|\mathbf{r}-\mathbf{r}'|^{-1}$ is the bare Coulomb potential.
Additional computational details are provided as Supporting Information.

\clearpage

\clearpage

\begin{acknowledgement}

The authors thank David Beljonne and Henning Sirringhaus for stimulating discussion about doping in organic semiconductors.
High performance computing resources from GENCI-TGCC (grant no. 2020-A0090910016) are acknowledged.
M.C. acknowledges PhD scholarship funding from Grenoble Quantum Engineering (GreQuE) and Fondation Nanosciences as well as the European Union’s Horizon 2020 research and innovation programme under the Marie Skłodowska-Curie grant agreement No 754303 (GreQue).

\end{acknowledgement}

\begin{suppinfo}

This will usually read something like: ``Experimental procedures and
characterization data for all new compounds. The class will
automatically add a sentence pointing to the information on-line:

\end{suppinfo}

\bibliography{doping}

\providecommand{\latin}[1]{#1}
\makeatletter
\providecommand{\doi}
  {\begingroup\let\do\@makeother\dospecials
  \catcode`\{=1 \catcode`\}=2 \doi@aux}
\providecommand{\doi@aux}[1]{\endgroup\texttt{#1}}
\makeatother
\providecommand*\mcitethebibliography{\thebibliography}
\csname @ifundefined\endcsname{endmcitethebibliography}
  {\let\endmcitethebibliography\endthebibliography}{}
\begin{mcitethebibliography}{46}
\providecommand*\natexlab[1]{#1}
\providecommand*\mciteSetBstSublistMode[1]{}
\providecommand*\mciteSetBstMaxWidthForm[2]{}
\providecommand*\mciteBstWouldAddEndPuncttrue
  {\def\EndOfBibitem{\unskip.}}
\providecommand*\mciteBstWouldAddEndPunctfalse
  {\let\EndOfBibitem\relax}
\providecommand*\mciteSetBstMidEndSepPunct[3]{}
\providecommand*\mciteSetBstSublistLabelBeginEnd[3]{}
\providecommand*\EndOfBibitem{}
\mciteSetBstSublistMode{f}
\mciteSetBstMaxWidthForm{subitem}{(\alph{mcitesubitemcount})}
\mciteSetBstSublistLabelBeginEnd
  {\mcitemaxwidthsubitemform\space}
  {\relax}
  {\relax}

\bibitem[Walzer \latin{et~al.}(2007)Walzer, Maennig, Pfeiffer, and Leo]{Wal07}
Walzer,~K.; Maennig,~B.; Pfeiffer,~M.; Leo,~K. Highly Efficient Organic Devices
  Based on Electrically Doped Transport Layers. \emph{Chem. Rev.}
  \textbf{2007}, \emph{107}, 1233--1271\relax
\mciteBstWouldAddEndPuncttrue
\mciteSetBstMidEndSepPunct{\mcitedefaultmidpunct}
{\mcitedefaultendpunct}{\mcitedefaultseppunct}\relax
\EndOfBibitem
\bibitem[Jacobs and Moulé(2017)Jacobs, and Moulé]{Jac17}
Jacobs,~I.; Moulé,~A. Controlling Molecular Doping in Organic Semiconductors.
  \emph{Adv. Mater.} \textbf{2017}, \emph{29}, 1703063\relax
\mciteBstWouldAddEndPuncttrue
\mciteSetBstMidEndSepPunct{\mcitedefaultmidpunct}
{\mcitedefaultendpunct}{\mcitedefaultseppunct}\relax
\EndOfBibitem
\bibitem[Salzmann \latin{et~al.}(2016)Salzmann, Heimel, Oehzelt, Winkler, and
  Koch]{Sal16}
Salzmann,~I.; Heimel,~G.; Oehzelt,~M.; Winkler,~S.; Koch,~N. Molecular
  Electrical Doping of Organic Semiconductors: Fundamental Mechanisms and
  Emerging Dopant Design Rules. \emph{Acc. Chem. Res.} \textbf{2016},
  \emph{49}, 370--378\relax
\mciteBstWouldAddEndPuncttrue
\mciteSetBstMidEndSepPunct{\mcitedefaultmidpunct}
{\mcitedefaultendpunct}{\mcitedefaultseppunct}\relax
\EndOfBibitem
\bibitem[Zhao \latin{et~al.}(2020)Zhao, Ding, Zou, Di, and Zhu]{Zhao20}
Zhao,~W.; Ding,~J.; Zou,~Y.; Di,~C.-a.; Zhu,~D. Chemical doping of organic
  semiconductors for thermoelectric applications. \emph{Chem. Soc. Rev.}
  \textbf{2020}, 7210--7228\relax
\mciteBstWouldAddEndPuncttrue
\mciteSetBstMidEndSepPunct{\mcitedefaultmidpunct}
{\mcitedefaultendpunct}{\mcitedefaultseppunct}\relax
\EndOfBibitem
\bibitem[Lin \latin{et~al.}(2017)Lin, Wegner, Lee, Fusella, Zhang, Moudgil,
  Rand, Barlow, Marder, Koch, and Kahn]{Lin17}
Lin,~X.; Wegner,~B.; Lee,~K.~M.; Fusella,~M.~A.; Zhang,~F.; Moudgil,~K.;
  Rand,~B.~P.; Barlow,~S.; Marder,~S.~R.; Koch,~N. \latin{et~al.}  Beating the
  thermodynamic limit with photo-activation of n-doping in organic
  semiconductors. \emph{Nat. Mater.} \textbf{2017}, \emph{16}, 1209--1215\relax
\mciteBstWouldAddEndPuncttrue
\mciteSetBstMidEndSepPunct{\mcitedefaultmidpunct}
{\mcitedefaultendpunct}{\mcitedefaultseppunct}\relax
\EndOfBibitem
\bibitem[Yamashita \latin{et~al.}(2019)Yamashita, Tsurumi, Ohno, Fujimoto,
  Kumagai, Kurosawa, Okamoto, Takeya, and Watanabe]{Yam19}
Yamashita,~Y.; Tsurumi,~J.; Ohno,~M.; Fujimoto,~R.; Kumagai,~S.; Kurosawa,~T.;
  Okamoto,~T.; Takeya,~J.; Watanabe,~S. Efficient molecular doping of polymeric
  semiconductors driven by anion exchange. \emph{Nature} \textbf{2019},
  \emph{572}, 634--638\relax
\mciteBstWouldAddEndPuncttrue
\mciteSetBstMidEndSepPunct{\mcitedefaultmidpunct}
{\mcitedefaultendpunct}{\mcitedefaultseppunct}\relax
\EndOfBibitem
\bibitem[Yurash \latin{et~al.}(2019)Yurash, Cao, Brus, Leifert, Wang, Dixon,
  Seifrid, Mansour, Lungwitz, Liu, Santiago, Graham, Koch, Bazan, and
  Nguyen]{Yur19}
Yurash,~B.; Cao,~D.~X.; Brus,~V.~V.; Leifert,~D.; Wang,~M.; Dixon,~A.;
  Seifrid,~M.; Mansour,~A.~E.; Lungwitz,~D.; Liu,~T. \latin{et~al.}  Towards
  understanding the doping mechanism of organic semiconductors by Lewis acids.
  \emph{Nat. Mater.} \textbf{2019}, \emph{18}, 1327--1334\relax
\mciteBstWouldAddEndPuncttrue
\mciteSetBstMidEndSepPunct{\mcitedefaultmidpunct}
{\mcitedefaultendpunct}{\mcitedefaultseppunct}\relax
\EndOfBibitem
\bibitem[Kang \latin{et~al.}(2016)Kang, Watanabe, Broch, Sepe, Brown,
  Nasrallah, Nikolka, Fei, Heeney, and Matsumoto]{Kan16}
Kang,~K.; Watanabe,~S.; Broch,~K.; Sepe,~A.; Brown,~A.; Nasrallah,~I.;
  Nikolka,~M.; Fei,~Z.; Heeney,~M.; Matsumoto,~D. 2D coherent charge transport
  in highly ordered conducting polymers doped by solid state diffusion.
  \emph{Nat. mater.} \textbf{2016}, \emph{15}, 896--902\relax
\mciteBstWouldAddEndPuncttrue
\mciteSetBstMidEndSepPunct{\mcitedefaultmidpunct}
{\mcitedefaultendpunct}{\mcitedefaultseppunct}\relax
\EndOfBibitem
\bibitem[Hase \latin{et~al.}(2018)Hase, O’Neill, Frisch, Opitz, Koch, and
  Salzmann]{Hase18}
Hase,~H.; O’Neill,~K.; Frisch,~J.; Opitz,~A.; Koch,~N.; Salzmann,~I.
  Unraveling the Microstructure of Molecularly Doped Poly(3-hexylthiophene) by
  Thermally Induced Dedoping. \emph{The Journal of Physical Chemistry C}
  \textbf{2018}, \emph{122}, 25893--25899\relax
\mciteBstWouldAddEndPuncttrue
\mciteSetBstMidEndSepPunct{\mcitedefaultmidpunct}
{\mcitedefaultendpunct}{\mcitedefaultseppunct}\relax
\EndOfBibitem
\bibitem[Jacobs \latin{et~al.}(2018)Jacobs, Cendra, Harrelson, Bedolla~Valdez,
  Faller, Salleo, and Moulé]{Jac18}
Jacobs,~I.~E.; Cendra,~C.; Harrelson,~T.~F.; Bedolla~Valdez,~Z.~I.; Faller,~R.;
  Salleo,~A.; Moulé,~A.~J. Polymorphism controls the degree of charge transfer
  in a molecularly doped semiconducting polymer. \emph{Mater. Horiz.}
  \textbf{2018}, \emph{5}, 655--660\relax
\mciteBstWouldAddEndPuncttrue
\mciteSetBstMidEndSepPunct{\mcitedefaultmidpunct}
{\mcitedefaultendpunct}{\mcitedefaultseppunct}\relax
\EndOfBibitem
\bibitem[Warren \latin{et~al.}(2019)Warren, Privitera, Kaienburg, Lauritzen,
  Thimm, Nelson, and Riede]{War19}
Warren,~R.; Privitera,~A.; Kaienburg,~P.; Lauritzen,~A.~E.; Thimm,~O.;
  Nelson,~J.; Riede,~M.~K. Controlling energy levels and Fermi level en route
  to fully tailored energetics in organic semiconductors. \emph{Nat. Commun.}
  \textbf{2019}, \emph{10}\relax
\mciteBstWouldAddEndPuncttrue
\mciteSetBstMidEndSepPunct{\mcitedefaultmidpunct}
{\mcitedefaultendpunct}{\mcitedefaultseppunct}\relax
\EndOfBibitem
\bibitem[Kleemann \latin{et~al.}(2012)Kleemann, Schuenemann, Zakhidov, Riede,
  L{\"u}ssem, and Leo]{Kle12}
Kleemann,~H.; Schuenemann,~C.; Zakhidov,~A.~A.; Riede,~M.; L{\"u}ssem,~B.;
  Leo,~K. Structural phase transition in pentacene caused by molecular doping
  and its effect on charge carrier mobility. \emph{Org. Electron.}
  \textbf{2012}, \emph{13}, 58 -- 65\relax
\mciteBstWouldAddEndPuncttrue
\mciteSetBstMidEndSepPunct{\mcitedefaultmidpunct}
{\mcitedefaultendpunct}{\mcitedefaultseppunct}\relax
\EndOfBibitem
\bibitem[Mendez \latin{et~al.}(2015)Mendez, Heimel, Winkler, Frisch, Opitz,
  Sauer, Wegner, Oehzelt, Rothel, Duhm, Tobbens, Koch, and Salzmann]{Men15}
Mendez,~H.; Heimel,~G.; Winkler,~S.; Frisch,~J.; Opitz,~A.; Sauer,~K.;
  Wegner,~B.; Oehzelt,~M.; Rothel,~C.; Duhm,~S. \latin{et~al.}  Charge-transfer
  crystallites as molecular electrical dopants. \emph{Nat. Commun.}
  \textbf{2015}, \emph{6}, 8560\relax
\mciteBstWouldAddEndPuncttrue
\mciteSetBstMidEndSepPunct{\mcitedefaultmidpunct}
{\mcitedefaultendpunct}{\mcitedefaultseppunct}\relax
\EndOfBibitem
\bibitem[Pingel and Neher(2013)Pingel, and Neher]{Pin13}
Pingel,~P.; Neher,~D. Comprehensive Picture of p-Type Doping of P3HT With the
  Molecular Acceptor F$_{4}$TCNQ. \emph{Physical Review B} \textbf{2013},
  \emph{87}, 115209\relax
\mciteBstWouldAddEndPuncttrue
\mciteSetBstMidEndSepPunct{\mcitedefaultmidpunct}
{\mcitedefaultendpunct}{\mcitedefaultseppunct}\relax
\EndOfBibitem
\bibitem[Maennig \latin{et~al.}(2001)Maennig, Pfeiffer, Nollau, Zhou, Leo, and
  Simon]{Mae01}
Maennig,~B.; Pfeiffer,~M.; Nollau,~A.; Zhou,~X.; Leo,~K.; Simon,~P. Controlled
  p-type doping of polycrystalline and amorphous organic layers:
  Self-consistent description of conductivity and field-effect mobility by a
  microscopic percolation model. \emph{Phys. Rev. B} \textbf{2001}, \emph{64},
  195208\relax
\mciteBstWouldAddEndPuncttrue
\mciteSetBstMidEndSepPunct{\mcitedefaultmidpunct}
{\mcitedefaultendpunct}{\mcitedefaultseppunct}\relax
\EndOfBibitem
\bibitem[Shen \latin{et~al.}(2003)Shen, Diest, Wong, Hsieh, Dunlap, and
  Malliaras]{She03}
Shen,~Y.; Diest,~K.; Wong,~M.~H.; Hsieh,~B.~R.; Dunlap,~D.~H.; Malliaras,~G.~G.
  Charge transport in doped organic semiconductors. \emph{Phys. Rev. B}
  \textbf{2003}, \emph{68}, 081204\relax
\mciteBstWouldAddEndPuncttrue
\mciteSetBstMidEndSepPunct{\mcitedefaultmidpunct}
{\mcitedefaultendpunct}{\mcitedefaultseppunct}\relax
\EndOfBibitem
\bibitem[Olthof \latin{et~al.}(2012)Olthof, Mehraeen, Mohapatra, Barlow,
  Coropceanu, Br\'edas, Marder, and Kahn]{Olt12}
Olthof,~S.; Mehraeen,~S.; Mohapatra,~S.~K.; Barlow,~S.; Coropceanu,~V.;
  Br\'edas,~J.-L.; Marder,~S.~R.; Kahn,~A. Ultralow Doping in Organic
  Semiconductors: Evidence of Trap Filling. \emph{Phys. Rev. Lett.}
  \textbf{2012}, \emph{109}, 176601\relax
\mciteBstWouldAddEndPuncttrue
\mciteSetBstMidEndSepPunct{\mcitedefaultmidpunct}
{\mcitedefaultendpunct}{\mcitedefaultseppunct}\relax
\EndOfBibitem
\bibitem[Schwarze \latin{et~al.}(2019)Schwarze, Gaul, Scholz, Bussolotti,
  Hofacker, Schellhammer, Nell, Naab, Bao, Spoltore, Vandewal, Widmer, Kera,
  Ueno, Ortmann, and Leo]{Schwa19}
Schwarze,~M.; Gaul,~C.; Scholz,~R.; Bussolotti,~F.; Hofacker,~A.;
  Schellhammer,~K.~S.; Nell,~B.; Naab,~B.~D.; Bao,~Z.; Spoltore,~D.
  \latin{et~al.}  Molecular parameters responsible for thermally activated
  transport in doped organic semiconductors. \emph{Nat. Mater.} \textbf{2019},
  \emph{18}, 242\relax
\mciteBstWouldAddEndPuncttrue
\mciteSetBstMidEndSepPunct{\mcitedefaultmidpunct}
{\mcitedefaultendpunct}{\mcitedefaultseppunct}\relax
\EndOfBibitem
\bibitem[Gaul \latin{et~al.}(2018)Gaul, Hutsch, Schwarze, Schellhammer,
  Bussolotti, Kera, Cuniberti, Leo, and Ortmann]{Gau18}
Gaul,~C.; Hutsch,~S.; Schwarze,~M.; Schellhammer,~K.~S.; Bussolotti,~F.;
  Kera,~S.; Cuniberti,~G.; Leo,~K.; Ortmann,~F. Insight into doping efficiency
  of organic semiconductors from the analysis of the density of states in
  n-doped C60 and ZnPc. \emph{Nat. Mater.} \textbf{2018}, \emph{17}, 439\relax
\mciteBstWouldAddEndPuncttrue
\mciteSetBstMidEndSepPunct{\mcitedefaultmidpunct}
{\mcitedefaultendpunct}{\mcitedefaultseppunct}\relax
\EndOfBibitem
\bibitem[Li \latin{et~al.}(2017)Li, D'Avino, Pershin, Jacquemin, Duchemin,
  Beljonne, and Blase]{Li17}
Li,~J.; D'Avino,~G.; Pershin,~A.; Jacquemin,~D.; Duchemin,~I.; Beljonne,~D.;
  Blase,~X. Correlated electron-hole mechanism for molecular doping in organic
  semiconductors. \emph{Phys. Rev. Mater.} \textbf{2017}, \emph{1},
  025602\relax
\mciteBstWouldAddEndPuncttrue
\mciteSetBstMidEndSepPunct{\mcitedefaultmidpunct}
{\mcitedefaultendpunct}{\mcitedefaultseppunct}\relax
\EndOfBibitem
\bibitem[Li \latin{et~al.}(2019)Li, Duchemin, Roscioni, Friederich, Anderson,
  Como, Kociok-Köhn, Wenzel, Zannoni, Beljonne, Blase, and D'Avino]{Li19}
Li,~J.; Duchemin,~I.; Roscioni,~O.~M.; Friederich,~P.; Anderson,~M.;
  Como,~E.~D.; Kociok-Köhn,~G.; Wenzel,~W.; Zannoni,~C.; Beljonne,~D.
  \latin{et~al.}  Host dependence of the electron affinity of molecular
  dopants. \emph{Mater. Horiz.} \textbf{2019}, \emph{6}, 107\relax
\mciteBstWouldAddEndPuncttrue
\mciteSetBstMidEndSepPunct{\mcitedefaultmidpunct}
{\mcitedefaultendpunct}{\mcitedefaultseppunct}\relax
\EndOfBibitem
\bibitem[Privitera \latin{et~al.}(2020)Privitera, Londi, Riede, D'Avino, and
  Beljonne]{Pri20}
Privitera,~A.; Londi,~G.; Riede,~M.; D'Avino,~G.; Beljonne,~D. Molecular
  Quadrupole Moments Promote Ground-State Charge Generation in Doped Organic
  Semiconductors. \emph{Adv. Funct. Mater.} \textbf{2020}, \emph{30},
  2004600\relax
\mciteBstWouldAddEndPuncttrue
\mciteSetBstMidEndSepPunct{\mcitedefaultmidpunct}
{\mcitedefaultendpunct}{\mcitedefaultseppunct}\relax
\EndOfBibitem
\bibitem[Zhang and Kahn(2018)Zhang, and Kahn]{Zha18}
Zhang,~F.; Kahn,~A. Investigation of the High Electron Affinity Molecular
  Dopant F6-TCNNQ for Hole-Transport Materials. \emph{Advanced Functional
  Materials} \textbf{2018}, \emph{28}, 1703780\relax
\mciteBstWouldAddEndPuncttrue
\mciteSetBstMidEndSepPunct{\mcitedefaultmidpunct}
{\mcitedefaultendpunct}{\mcitedefaultseppunct}\relax
\EndOfBibitem
\bibitem[Jacobs \latin{et~al.}(2021)Jacobs, D'Avino, Lin, Lemaur, Huang, Ren,
  Simatos, Wood, Chen, Harrelson, Mustafa, O'Keefe, Spalek, Tjhe, Statz, Lai,
  Finn, Neal, Strzalka, Nielsen, Lee, Barlow, Marder, McCulloch, Fratini,
  Beljonne, and Sirringhaus]{Jac21}
Jacobs,~I.~E.; D'Avino,~G.; Lin,~Y.; Lemaur,~V.; Huang,~Y.; Ren,~X.;
  Simatos,~D.; Wood,~W.; Chen,~C.; Harrelson,~T. \latin{et~al.}  Ion-exchange
  doped polymers at the degenerate limit: what limits conductivity at 100\%
  doping efficiency? \emph{arXiv} \textbf{2021}, 2101.01714\relax
\mciteBstWouldAddEndPuncttrue
\mciteSetBstMidEndSepPunct{\mcitedefaultmidpunct}
{\mcitedefaultendpunct}{\mcitedefaultseppunct}\relax
\EndOfBibitem
\bibitem[Tietze \latin{et~al.}(2015)Tietze, Pahner, Schmidt, Leo, and
  Lüssem]{Tie15}
Tietze,~M.~L.; Pahner,~P.; Schmidt,~K.; Leo,~K.; Lüssem,~B. Doped Organic
  Semiconductors: Trap-Filling, Impurity Saturation, and Reserve Regimes.
  \emph{Adv. Funct. Mater.} \textbf{2015}, \emph{25}, 2701--2707\relax
\mciteBstWouldAddEndPuncttrue
\mciteSetBstMidEndSepPunct{\mcitedefaultmidpunct}
{\mcitedefaultendpunct}{\mcitedefaultseppunct}\relax
\EndOfBibitem
\bibitem[Salzmann \latin{et~al.}(2012)Salzmann, Heimel, Duhm, Oehzelt, Pingel,
  George, Schnegg, Lips, Blum, Vollmer, and Koch]{Sal12}
Salzmann,~I.; Heimel,~G.; Duhm,~S.; Oehzelt,~M.; Pingel,~P.; George,~B.~M.;
  Schnegg,~A.; Lips,~K.; Blum,~R. l.-P.; Vollmer,~A. \latin{et~al.}
  Intermolecular Hybridization Governs Molecular Electrical Doping. \emph{Phys.
  Rev. Lett.} \textbf{2012}, \emph{108}, 035502\relax
\mciteBstWouldAddEndPuncttrue
\mciteSetBstMidEndSepPunct{\mcitedefaultmidpunct}
{\mcitedefaultendpunct}{\mcitedefaultseppunct}\relax
\EndOfBibitem
\bibitem[Valencia and Cocchi(2019)Valencia, and Cocchi]{Val19}
Valencia,~A.~M.; Cocchi,~C. Electronic and Optical Properties of
  Oligothiophene-F4TCNQ Charge-Transfer Complexes: The Role of the Donor
  Conjugation Length. \emph{J. Phys. Chem. C} \textbf{2019}, \emph{123},
  9617--9623\relax
\mciteBstWouldAddEndPuncttrue
\mciteSetBstMidEndSepPunct{\mcitedefaultmidpunct}
{\mcitedefaultendpunct}{\mcitedefaultseppunct}\relax
\EndOfBibitem
\bibitem[Ha and Kahn(2009)Ha, and Kahn]{Ha09}
Ha,~S.~D.; Kahn,~A. Isolated molecular dopants in pentacene observed by
  scanning tunneling microscopy. \emph{Phys. Rev. B} \textbf{2009}, \emph{80},
  195410\relax
\mciteBstWouldAddEndPuncttrue
\mciteSetBstMidEndSepPunct{\mcitedefaultmidpunct}
{\mcitedefaultendpunct}{\mcitedefaultseppunct}\relax
\EndOfBibitem
\bibitem[Tietze \latin{et~al.}(2018)Tietze, Benduhn, Pahner, Nell, Schwarze,
  Kleemann, Krammer, Zojer, Vandewal, and Leo]{Tie18}
Tietze,~M.~L.; Benduhn,~J.; Pahner,~P.; Nell,~B.; Schwarze,~M.; Kleemann,~H.;
  Krammer,~M.; Zojer,~K.; Vandewal,~K.; Leo,~K. Elementary steps in electrical
  doping of organic semiconductors. \emph{Nat. Commun.} \textbf{2018},
  \emph{9}, 1182\relax
\mciteBstWouldAddEndPuncttrue
\mciteSetBstMidEndSepPunct{\mcitedefaultmidpunct}
{\mcitedefaultendpunct}{\mcitedefaultseppunct}\relax
\EndOfBibitem
\bibitem[Fediai \latin{et~al.}(2019)Fediai, Symalla, Friederich, and
  Wenzel]{Fed19}
Fediai,~A.; Symalla,~F.; Friederich,~P.; Wenzel,~W. Disorder compensation
  controls doping efficiency in organic semiconductors. \emph{Nat. Commun.}
  \textbf{2019}, \emph{10}, 4547\relax
\mciteBstWouldAddEndPuncttrue
\mciteSetBstMidEndSepPunct{\mcitedefaultmidpunct}
{\mcitedefaultendpunct}{\mcitedefaultseppunct}\relax
\EndOfBibitem
\bibitem[Fediai \latin{et~al.}(2020)Fediai, Emering, Symalla, and
  Wenzel]{Fed20}
Fediai,~A.; Emering,~A.; Symalla,~F.; Wenzel,~W. Disorder-driven doping
  activation in organic semiconductors. \emph{Phys. Chem. Chem. Phys.}
  \textbf{2020}, \emph{22}, 10256--10264\relax
\mciteBstWouldAddEndPuncttrue
\mciteSetBstMidEndSepPunct{\mcitedefaultmidpunct}
{\mcitedefaultendpunct}{\mcitedefaultseppunct}\relax
\EndOfBibitem
\bibitem[Koopmans \latin{et~al.}(2020)Koopmans, Leiviskä, Liu, Dong, Qiu,
  Hummelen, Portale, Heiber, and Koster]{Koo20}
Koopmans,~M.; Leiviskä,~M. A.~T.; Liu,~J.; Dong,~J.; Qiu,~L.; Hummelen,~J.~C.;
  Portale,~G.; Heiber,~M.~C.; Koster,~L. J.~A. Electrical Conductivity of Doped
  Organic Semiconductors Limited by Carrier–Carrier Interactions. \emph{ACS
  Applied Materials \& Interfaces} \textbf{2020}, \emph{12}, 56222--56230\relax
\mciteBstWouldAddEndPuncttrue
\mciteSetBstMidEndSepPunct{\mcitedefaultmidpunct}
{\mcitedefaultendpunct}{\mcitedefaultseppunct}\relax
\EndOfBibitem
\bibitem[Li \latin{et~al.}(2018)Li, D'Avino, Duchemin, Beljonne, and
  Blase]{Li18}
Li,~J.; D'Avino,~G.; Duchemin,~I.; Beljonne,~D.; Blase,~X. Accurate description
  of charged excitations in molecular solids from embedded many-body
  perturbation theory. \emph{Phys. Rev. B} \textbf{2018}, \emph{97},
  035108\relax
\mciteBstWouldAddEndPuncttrue
\mciteSetBstMidEndSepPunct{\mcitedefaultmidpunct}
{\mcitedefaultendpunct}{\mcitedefaultseppunct}\relax
\EndOfBibitem
\bibitem[D'Avino \latin{et~al.}(2014)D'Avino, Muccioli, Zannoni, Beljonne, and
  Soos]{Dav14}
D'Avino,~G.; Muccioli,~L.; Zannoni,~C.; Beljonne,~D.; Soos,~Z.~G. Electronic
  Polarization in Organic Crystals: A Comparative Study of Induced Dipoles and
  Intramolecular Charge Redistribution Schemes. \emph{J. Chem. Theory Comput.}
  \textbf{2014}, \emph{10}, 4959\relax
\mciteBstWouldAddEndPuncttrue
\mciteSetBstMidEndSepPunct{\mcitedefaultmidpunct}
{\mcitedefaultendpunct}{\mcitedefaultseppunct}\relax
\EndOfBibitem
\bibitem[Mueller(1935)]{mueller_PhysRev35}
Mueller,~H. Theory of the Photoelastic Effect of Cubic Crystals. \emph{Phys.
  Rev.} \textbf{1935}, \emph{47}, 947 -- 957\relax
\mciteBstWouldAddEndPuncttrue
\mciteSetBstMidEndSepPunct{\mcitedefaultmidpunct}
{\mcitedefaultendpunct}{\mcitedefaultseppunct}\relax
\EndOfBibitem
\bibitem[Colpa(1971)]{Colpa_Physica71A}
Colpa,~J. Dipole fields and electric-field gradients in their dependence on the
  macroscopic and microscopic crystal parameters for orthorhombic and hexagonal
  lattices. I. \emph{Physica} \textbf{1971}, \emph{56}, 185 -- 204\relax
\mciteBstWouldAddEndPuncttrue
\mciteSetBstMidEndSepPunct{\mcitedefaultmidpunct}
{\mcitedefaultendpunct}{\mcitedefaultseppunct}\relax
\EndOfBibitem
\bibitem[Purvis and Taylor(1982)Purvis, and Taylor]{Purvis_taylor_PRB82}
Purvis,~C.~K.; Taylor,~P.~L. Dipole-field sums and Lorentz factors for
  orthorhombic lattices, and implications for polarizable molecules.
  \emph{Phys. Rev. B} \textbf{1982}, \emph{26}, 4547 -- 4563\relax
\mciteBstWouldAddEndPuncttrue
\mciteSetBstMidEndSepPunct{\mcitedefaultmidpunct}
{\mcitedefaultendpunct}{\mcitedefaultseppunct}\relax
\EndOfBibitem
\bibitem[Vanzo \latin{et~al.}(2015)Vanzo, Topham, and Soos]{Van15}
Vanzo,~D.; Topham,~B.~J.; Soos,~Z.~G. Dipole-Field Sums, Lorentz Factors, and
  Dielectric Properties of Organic Molecular Films Modeled as Crystalline
  Arrays of Polarizable Points. \emph{Adv. Funct. Mater.} \textbf{2015},
  \emph{25}, 2004--2012\relax
\mciteBstWouldAddEndPuncttrue
\mciteSetBstMidEndSepPunct{\mcitedefaultmidpunct}
{\mcitedefaultendpunct}{\mcitedefaultseppunct}\relax
\EndOfBibitem
\bibitem[Herzfeld(1927)]{Her27}
Herzfeld,~K.~F. On Atomic Properties which make an Element a Metal. \emph{Phys.
  Rev.} \textbf{1927}, \emph{29}, 701--705\relax
\mciteBstWouldAddEndPuncttrue
\mciteSetBstMidEndSepPunct{\mcitedefaultmidpunct}
{\mcitedefaultendpunct}{\mcitedefaultseppunct}\relax
\EndOfBibitem
\bibitem[Tsiper and Soos(2001)Tsiper, and Soos]{Tsi01}
Tsiper,~E.~V.; Soos,~Z.~G. Charge redistribution and polarization energy of
  organic molecular crystals. \emph{Phys. Rev. B} \textbf{2001}, \emph{64},
  195124\relax
\mciteBstWouldAddEndPuncttrue
\mciteSetBstMidEndSepPunct{\mcitedefaultmidpunct}
{\mcitedefaultendpunct}{\mcitedefaultseppunct}\relax
\EndOfBibitem
\bibitem[D’Avino \latin{et~al.}(2016)D’Avino, Vanzo, and Soos]{Dav16}
D’Avino,~G.; Vanzo,~D.; Soos,~Z.~G. Dielectric properties of crystalline
  organic molecular films in the limit of zero overlap. \emph{J. Chem. Phys.}
  \textbf{2016}, \emph{144}, 034702\relax
\mciteBstWouldAddEndPuncttrue
\mciteSetBstMidEndSepPunct{\mcitedefaultmidpunct}
{\mcitedefaultendpunct}{\mcitedefaultseppunct}\relax
\EndOfBibitem
\bibitem[Gregg(2011)]{Gregg11}
Gregg,~B.~A. Entropy of Charge Separation in Organic Photovoltaic Cells: The
  Benefit of Higher Dimensionality. \emph{J. Phys. Chem. Lett.} \textbf{2011},
  \emph{2}, 3013--3015\relax
\mciteBstWouldAddEndPuncttrue
\mciteSetBstMidEndSepPunct{\mcitedefaultmidpunct}
{\mcitedefaultendpunct}{\mcitedefaultseppunct}\relax
\EndOfBibitem
\bibitem[Castellan and Seitz(1951)Castellan, and Seitz]{Cas51}
Castellan,~G.~M.; Seitz,~F. \emph{Semiconducting Materials}; Butterworths:
  London, 1951\relax
\mciteBstWouldAddEndPuncttrue
\mciteSetBstMidEndSepPunct{\mcitedefaultmidpunct}
{\mcitedefaultendpunct}{\mcitedefaultseppunct}\relax
\EndOfBibitem
\bibitem[Dhar and Marshak(1985)Dhar, and Marshak]{Dhar85}
Dhar,~S.; Marshak,~A.~H. Static dielectric constant of heavily doped
  semiconductors. \emph{Solid-State Electron.} \textbf{1985}, \emph{28},
  763--766\relax
\mciteBstWouldAddEndPuncttrue
\mciteSetBstMidEndSepPunct{\mcitedefaultmidpunct}
{\mcitedefaultendpunct}{\mcitedefaultseppunct}\relax
\EndOfBibitem
\bibitem[Mott and Kaveh(1985)Mott, and Kaveh]{Mott85}
Mott,~N.; Kaveh,~M. Metal-insulator transitions in non-crystalline systems.
  \emph{Adv. Phys.} \textbf{1985}, \emph{34}, 329--401\relax
\mciteBstWouldAddEndPuncttrue
\mciteSetBstMidEndSepPunct{\mcitedefaultmidpunct}
{\mcitedefaultendpunct}{\mcitedefaultseppunct}\relax
\EndOfBibitem
\end{mcitethebibliography}

\end{document}